\documentclass[conference,onecolumn]{IEEEtran}
\ifCLASSINFOpdf
\else
   \usepackage[dvips]{graphicx}
\fi
\usepackage{url}
\usepackage{graphicx}
\usepackage{amsmath}
\usepackage{comment}
\usepackage[utf8]{inputenc}
\usepackage[english]{babel}
\usepackage{color}
\usepackage{framed}

\usepackage{amsmath,amssymb,amsfonts}
\newenvironment{psmallmatrix}
  {\left(\begin{smallmatrix}}
  {\end{smallmatrix}\right)}

\usepackage{graphicx}
\usepackage{upgreek}
\usepackage{textcomp}

\usepackage{url}

\usepackage{cite}

\usepackage[dvipsnames]{xcolor}
\usepackage{amsmath}
\usepackage{amsthm}
\usepackage{dsfont}
\usepackage{thmtools}
\usepackage{amssymb}

\def\bGamma{\boldsymbol{\Gamma}}

\def\mba{\mathbf{a}}

\def\mbf{\mathbf{f}}

\def\mbs{\mathbf{s}}

\def\mbx{\mathbf{x}}
\def\mby{\mathbf{y}}

\def\mbB{\mathbf{B}}

\def\mbG{\mathbf{G}}

\def\mbI{\mathbf{I}}
\def\mbJ{\mathbf{J}}

\def\mbP{\mathbf{P}}

\def\mbR{\mathbf{R}}

\def\mbU{\mathbf{U}}

\def\mbX{\mathbf{X}}

\def\vec#1{\mathrm{vec}\left(#1\right)}

\theoremstyle{definition}

\usepackage{bbold}
\usepackage{ifpdf}
\usepackage{times}
\usepackage{layout}
\usepackage{float}
\usepackage{afterpage}
\usepackage{amsmath}
\usepackage{amstext}
\usepackage{amssymb,bm}
\usepackage{latexsym}
\usepackage{color}
\usepackage{flexisym}
\usepackage{kantlipsum}
\usepackage{graphicx}
\usepackage{amsmath}
\usepackage{amsthm}
\usepackage{graphicx}

\usepackage[caption=false,font=footnotesize]{subfig}

\usepackage{booktabs}
\usepackage{multicol}

\usepackage{enumitem}
\newenvironment{bsmallmatrix}
  {\left[\begin{smallmatrix}}
  {\end{smallmatrix}\right]}

\usepackage[switch]{lineno}
\usepackage[named]{algo}
\usepackage[noend]{algpseudocode}
\usepackage{algorithm}

\makeatletter
\newcommand*{\rom}[1]{\expandafter\@slowromancap\romannumeral #1@}
\makeatother

\begin{document}
\setlength{\abovedisplayskip}{3pt}
\setlength{\belowdisplayskip}{3pt}

\title{Near-Field Low-WISL Unimodular Waveform Design for Terahertz Automotive Radar }

\author{\IEEEauthorblockN{Arian Eamaz$^{\dag,1}$, Farhang Yeganegi$^{\dag,^2}$, Kumar Vijay Mishra$^{\ddag,^3}$ and Mojtaba Soltanalian$^{\dag,^4}$}
\IEEEauthorblockA{{$^{\dag}$University of Illinois, Chicago, IL 60607, USA} \\
$^{\ddag}$United States DEVCOM Army Research Laboratory, Adelphi, MD 20783 USA\\
email: \{$^1$aeamaz2, $^2$fyegan2, $^4$msol\}@uic.edu, $^3$kvm@ieee.org}
}


\maketitle

\begin{abstract}
Conventional sensing applications rely on electromagnetic far-field channel models with plane wave propagation. However, recent ultra-short-range automotive radar applications at upper millimeter-wave or low terahertz (THz) frequencies envisage operation in the \emph{near-field} region, where the wavefront is spherical. Unlike far-field, the near-field beampattern is dependent on both range and angle, thus requiring a different approach to waveform design. For the first time in the literature, we adopt the \emph{beampattern matching} approach to design unimodular waveforms for THz automotive radars with low weighted integrated sidelobe levels (WISL). We formulate this problem as a unimodular bi-quadratic matrix program, and solve its constituent quadratic sub-problems using our cyclic power method-like iterations (CyPMLI) algorithm. Numerical experiments demonstrate that the CyPMLI approach yields the required beampattern with low autocorrelation levels.
\end{abstract}


\setlength{\abovedisplayskip}{3pt}
\setlength{\belowdisplayskip}{3pt}

\section{Introduction}
The shape of the propagating wavefront varies depending on the observation distance \cite{gustafsson2006characterization,glazunov2009spherical,glazunov2010physical}. 
Accordingly, three distinct regions have been identified
: near-field, Fresnel, and far-field (Fraunhofer) regions. The reciprocity of channels usually implies that these regions may be viewed as such from both transmitter and receiver perspectives \cite{selvan2017fraunhofer}. In the near-field, amplitude variations over the antenna aperture are noticeable \cite{friedlander2019localization}. In contrast, these variations are negligible in the Fresnel region, but phase variations still occur because of the signal's wavelength. In the far-field, both amplitude and phase variations are negligible; the amplitude (phase) depends on only the propagation distance (signal's incident angle) and the wavefront is approximated as locally planar. This leads to a linear propagation model via Fourier theory. 

Radar systems at lower sub-6 GHz frequencies, where the antenna array is electromagnetically smaller than the operating wavelength, rely on far-field plane-wave models \cite{yanik2020development}. However, with the advent of automotive radar applications at millimeter-wave and terahertz (THz) frequencies \cite{elbir2022terahertzband,elbir2021terahertz} that employ electrically large arrays, the far-field assumption breaks down for short-range operation \cite{sarieddeen2021overview,mishra2023signal}. At such ranges, the wavefront becomes spherical in the near-field  \cite{gustafsson2006characterization,glazunov2009spherical,glazunov2010physical}, thereby requiring the use of Weyl's decomposition \cite{weyl1919ausbreitung} of the spherical wave into several plane waves \cite{yanik2020development,meng2021accelerated}. This manifests itself in the array beampattern becoming a function of both angle and range \cite{elbir2023nba}.

Some far-field applications also exhibit range-dependent beampattern such as frequency diverse array (FDA) radars \cite{lv2022co,lv2022clutter}, wherein linear frequency offsets in the carrier frequency across array elements results in a range-dependent beampattern without a spherical wavefront. Similar complex patterns are observed in quantum Rydberg arrays \cite{vouras2023overview,vouras2023phase}. In this paper, we focus on near-field THz-band automotive radars \cite{bhattacharjee2023multi} that require consideration of range-dependent beampattern in system design \cite{mishra2023signal,elbir2022twenty,elbir2023spatial}. THz-band automotive radars have attracted considerable research interest in recent years because of their potential for a near-optical resolution \cite{bhattacharjee2023multi,mishra2023signal}. While the literature indicates a maximum of $200$ m range is possible \cite{norouzian2019rain} for THz automotive radars, most applications envisage their operation to 10-20 m \cite{bhattacharjee2023multi}. 

Prior works on THz automotive radar waveform design have included distance-dependent channel models \cite{bhattacharjee2023multi} and large arrays \cite{elbir2021terahertz} but ignored near-field range-dependent beampattern. Contrary to these works, we include near-field effects in our waveform design formulations. In particular, we focus on designing transmit signals with low correlation levels under the constraint of unimodularity \cite{stoica2009new,soltanalian2014designing,mohamadi2017arima}. The upshot of this approach is the minimal peak-to-average-ratio (PAR) 
and avoiding gain non-linearities with low-cost amplifiers \cite{cheng2017constant,blunt2014polyphase,aubry2015optimizing}. Automotive radars often employ multiple-input multiple-output (MIMO) arrays to improve resolution without using many antennas \cite{mishra2023signal}. In this case, the design problem requires obtaining a set of mutually (quasi-)orthogonal waveforms via minimization of low integrated sidelobe level (ISL) or weighted ISL (WISL) \cite{he2012waveform,li2022unimodular,alaee2022signal} thereby leading to improved target extraction \cite{aubry2016robust}, resolution \cite{li2015ambiguity}, and robustness \cite{he2012waveform}. 

We approach the near-field waveform design by adopting the beampattern matching approach \cite{stoica2009new,soltanalian2014single,mohamadi2023fussl}. The WISL metric for beampattern matching leads to a \emph{unimodular quartic matrix programming} (UQMP). We then formulate the near-field waveform design problem as a \emph{unimodular bi-quadratic matrix programming} (UBQMP). Here, a quartic-to-bi-quadratic transformation splits UQMP into two quadratic matrix subproblems  \cite{beck2007quadratic} that we solve using a low-complexity cyclic power method-like iterations (CyPMLI) algorithm \cite{eamaz2023cypmli,eamaz2023marli}. This is inspired by the power iteration method \cite{van1996matrix,soltanalian2014designing,soltanalian2013joint,mostofa2021deep}, which benefits from simple matrix-vector multiplications. Numerical experiments demonstrate that our proposed method achieves the desired beampattern while minimizing WISL. 
  
   
  Throughout this paper, we use bold lowercase and bold uppercase letters for vectors and matrices, respectively.  We represent a vector $\mathbf{x}\in\mathbb{C}^{N}$ in terms of its elements $\{x_{i}\}$ as $\mathbf{x}=[x_{i}]^{N}_{i=1}$. The $mn$-th element of the matrix $\mbB$ is $\left[\mbB\right]_{mn}$. The sets of complex and real numbers are $\mathbb{C}$ and $\mathbb{R}$, respectively;  $(\cdot)^{\top}$, $(\cdot)^{\ast}$and $(\cdot)^{\mathrm{H}}$ are the vector/matrix transpose, conjugate and the Hermitian transpose, respectively; trace of a matrix is  $\operatorname{Tr}(.)$; the function $\textrm{diag}(.)$ returns the diagonal elements of the input matrix. The Frobenius norm of a matrix $\mathbf{B}\in \mathbb{C}^{M\times N}$ is defined as $\|\mathbf{B}\|_{\mathrm{F}}=\sqrt{\sum^{M}_{r=1}\sum^{N}_{s=1}\left|b_{rs}\right|^{2}}$, where $b_{rs}$ is the $(r,s)$-th entry of $\mathbf{B}$. 
 The Hadamard (element-wise) and Kronecker products are $\odot$ and $\otimes$, respectively. The vectorized form of a matrix $\mbB$ is written as $\vec{\mbB}$. The $s$-dimensional all-ones vector, all-zeros vector, and the identity matrix of  size $s\times s$ are $\mathbf{1}_{s}$, $\mathbf{0}_{N}$, and $\mbI_s$, respectively. The real, imaginary, and angle/phase components of a complex number are $\operatorname{Re}(\cdot)$, $\operatorname{Im}(\cdot)$, and $\arg{(\cdot)}$, respectively.

\section{System Model}
\label{sec:system}

Consider a MIMO radar with $M$ linearly-spaced isotropic array elements, with the uniform inter-element spacing of $d$. The transmit antennas emit mutually orthogonal elements. The baseband signal transmitted by the $m$-th antenna is denoted by $x_{m}(t)$ with spectral support $\left[\frac{-B}{2},\frac{B}{2}\right]$, and continuous-time Fourier transform (CTFT),
\begin{equation}
\label{Neg7}
y_m(f)=\int_{-\infty}^{\infty} x_m(t) e^{-\textrm{j} 2 \pi f t} \,dt, \quad f \in\left[-\frac{B}{2}, \frac{B}{2}\right].
\end{equation}
The baseband signal is then upconverted for transmission, in the form $s_{m}(t)=x_{m}(t) e^{\textrm{j}2\pi f_{c}t}$, where $f_c$ denotes the carrier frequency.

The utilization of an extremely small array aperture that is electrically large compared to the wavelength leads to near-field interactions with targets in close proximity. When the transmission range is shorter than the Fraunhofer distance $F=\frac{2D^{2}}{\lambda}$, where $D=(M-1)d$ is the array aperture and $d=\frac{\lambda}{2}$ with $\lambda=\frac{c_{0}}{f}$ being the wavelength, the wavefront is spherical. At THz-band, the distance from the $k$-th target to the $l$-th path corresponding to the array origin, i.e., $p_{k,l}<F$ thereby requiring a near-field model \cite{elbir2022terahertz}. 

The near-field steering vector $\mba(\theta_{k,l}, p_{k,l})$ corresponding to physical direction-of-arrival (DoA) $\theta_{k,l}$ and range $p_{k,l}$, is 
\begin{equation}
\label{Neg1}
\mathbf{a}\left(\theta_{k, l}, r_{k, l}\right)=\frac{1}{\sqrt{M}}\left[e^{-\textrm{j} 2 \pi \frac{d}{\lambda} p_{k, l}^{(1)}}, \cdots, e^{-\mathrm{j} 2 \pi \frac{d}{\lambda} p_{k, l}^{(M)}}\right]^{\top},
\end{equation}
where $\theta_{k,l}=\sin{\phi_{k,l}}$, with $\phi_{k,l}\in\left[\frac{-\pi}{2},\frac{\pi}{2}\right]$ and $p^{(m)}_{k,l}$ is the distance between $k$-th target 
and the $m$-th antenna:
\begin{equation}
\label{Neg2}
p_{k, l}^{(m)}=\sqrt{p_{k, l}^2+2(m-1)^2 d^2-2 r_{k, l}(m-1) d \theta_{k, l}}.
\end{equation}
According to the Fresnel approximation 
\cite{gustafsson2006characterization,cui2022near}, we can approximate \eqref{Neg2} as
\begin{equation}
\label{Neg3}
p_{k, l}^{(m)} \approx p_{k, l}-(m-1) d \theta_{k, l}+(m-1)^2 d^2 \zeta_{k, l},
\end{equation}
where $\zeta_{k, l}=\frac{1-\theta_{k, l}^2}{2 p_{k, l}}$ is a function of both range and DoA. Substituting \eqref{Neg3} into \eqref{Neg1} gives 
\begin{equation}
\label{Neg4}
\mathbf{a}\left(\theta_{k, l}, p_{k, l}\right) \approx e^{-\textrm{j} 2 \pi \frac{f_c}{c_0} p_{k, l}} \tilde{\mathbf{a}}\left(\theta_{k, l}, p_{k, l}\right),
\end{equation}
where 
the $m$-th element of $\tilde{\mba}\in\mathbb{C}^{M}$ is  
$\left[\tilde{\mathbf{a}}\left(\theta_{k, l}, p_{k, l}\right)\right]_m=e^{\textrm{j} 2 \pi \frac{f_c}{c_0}\left((m-1) d \theta_{k, l}-(m-1)^2 d^2 \zeta_{k, l}\right)}$.


The (near-field) transmit signal at the location $\left(\theta_{k,l}, p_{k,l}\right)$ is 
\begin{equation}
\label{Neg6}
\begin{aligned}
z_{_{\theta_{k,l}, p_{k,l}}}(t) & =\sum_{m=1}^{M} s_m\left(t-\frac{d 
 p^{(m)}_{k,l}}{c_{0}}\right), \\
& =\sum_{m=1}^{M} x_m\left(t-\frac{d 
 p^{(m)}_{k,l}}{c_{0}}\right) e^{\textrm{j} 2 \pi f_c\left(t-\frac{d 
 p^{(m)}_{k,l}}{c_{0}}\right)}.
\end{aligned}
\end{equation}
Using inverse CTFT of \eqref{Neg7}, 
rewrite $z_{_{\theta_{k,l}, p_{k,l}}}(t)$ as
\begin{equation}
\label{Neg8}
z_{_{\theta_{k,l}, p_{k,l}}}(t)=\int_{-B / 2}^{B / 2} Y(\theta_{k,l},p_{k,l}, f) e^{\textrm{j} 2 \pi\left(f+f_c\right) t} \,df,
\end{equation}
where $Y(\theta_{k,l},p_{k,l}, f)=\sum_{m=1}^{M} y_m(f) e^{-\textrm{j} 2 \pi\left(f+f_c\right)\frac{d p^{(m)}_{k,l}}{c_{0}}}$.
As a result, the beampattern at location $\left\{\theta_{k,l},p_{k,l}\right\}$ and frequency $f+f_{c}$ is $ 
P(\theta_{k,l},p_{k,l}, f)=\left|Y(\theta_{k,l},p_{k,l}, f)\right|^2=\left|\boldsymbol{\upalpha}^{\mathrm{H}}(\theta_{k,l},p_{k,l}, f) \mathbf{y}(f)\right|^2$,
where $f \in\left[-\frac{B}{2}, \frac{B}{2}\right]$ and $\boldsymbol{\upalpha}$ is obtained based on the approximated near-field steering vector \eqref{Neg4}: 
\begin{equation}
\label{Neg11}
\boldsymbol{\upalpha}(\theta_{k,l},p_{k,l}, f)= e^{-\textrm{j}2\pi f} \mba^{\star}(\theta_{k,l},p_{k,l}),
\end{equation}
and $\mathbf{y}(f)=\left[\begin{array}{llll}
y_1(f) & y_1(f) & \cdots & y_{M}(f)
\end{array}\right]^\top$.
Sampling the signal $x_{m}(t)$  
at Nyquist interval $T_s=1/B$, we obtain $x_{m}(n)=x_{m}(nT_{s})$. The discrete Fourier transform (DFT) of $x_{m}(t)$ is 
\begin{equation}
\label{Neg13}
y_m(u)=\sum_{n=0}^{N-1} x_m(n) e^{-\textrm{j} 2 \pi \frac{n u}{N}}, ~ u \in \left\{0,1,\cdots,N-1\right\}.
\end{equation}
Define the vector $\mathbf{y}_u=\left[\begin{array}{llll}
y_0(u) & y_1(u) & \cdots & y_{M-1}(u)
\end{array}\right]^\top$. 

We assume that the DoAs and ranges/delays $\left\{\theta_{k,l}, p_{k,l}\right\}$ are aligned to grids $\left\{\theta_{k_{1}}\right\}^{K_{1}}_{k_{1}=1}$ and $\left\{p_{k_{2}}\right\}^{K_2}_{k_2=1}$, where  $\theta_{k_{1}}=\sin{\phi_{k_{1}}}$ with $\phi_{k_{1}}=\pi\left(\frac{k_1}{K_1}-\frac{1}{2}\right)$, $1<k_1<K_1$, 
and $p_{k_{2}}=\frac{k_2}{K_2}$, 
$1<k_2<K_2$. The grid size $K_1$ ($K_2$) is determined by the temporal (spatial) sampling rate.
The discretized $\boldsymbol{\upalpha}$ is 
\begin{equation}
\label{Neg14}
\boldsymbol{\upalpha}_{_{k_{1}, k_{2}, u}} = \boldsymbol{\upalpha}\left(\theta_{k_{1}},p_{k_{2}},\frac{u}{N T_{s}}\right).
\end{equation}
The discretized beampattern becomes
\begin{equation}
\label{Neg16}
P_{_{k_{1}, k_{2}, u}}=\left| \boldsymbol{\upalpha}_{_{k_{1}, k_{2},u}}^{\mathrm{H}} \mathbf{y}_u\right|^2.
\end{equation}

Our goal is to design waveform $\mbX=\left[\mbx_{1},\cdots,\mbx_{M}\right]\in \mathbb{C}^{M\times N}$ that focuses the beam in a desired direction. 

\section{Problem Formulation}
\label{sec:problem}
A two-stage algorithm for far-field wideband MIMO waveform design was suggested in \cite{li2007beampattern} based on the Gerchberg-Saxton algorithm \cite{gerchberg1972practical}. The key idea here is to obtain a complex-valued waveform in the spectral domain such that $\mby_{u}$ matches the magnitude of the desired beampattern as in \eqref{Neg16}. Related techniques also include phase-retrieval-based waveform design \cite{pinilla2022phase,pinilla2021wavemax}. 
We address the near-field version of this problem without resorting to phase retrieval methods. 

\subsection{Beampattern Matching Formulation}
\label{BMF}
Denote the desired beampattern by $\left\{\widehat{P}{_{k_{1}, k_{2}, u}}\right\}$ and $\Omega^{N}$ as the set of complex unimodular sequences
\begin{equation}
\label{eq:1}
\Omega^{N} = \left\{\mbs \in \mathbb{C}^{N}| 
s(l)=e^{\textrm{j}\omega_{l}}, \omega_l \in [0,2\pi),~0\leq l\leq N-1\right\}.
\end{equation}
The beampattern matching optimization problem is \cite{he2009designing},
\begin{equation}
\label{Neg18}
\begin{aligned}
&\underset{\mbx_m\in\Omega^{N}}{\textrm{minimize}} \sum_{k_{1}=1}^{K_{1}} \sum_{k_{2}=1}^{K_{2}} \sum_{u=0}^{N-1}\left[\widehat{P}{_{k_{1}, k_{2}, u}}-\left|\boldsymbol{\upalpha}_{_{k_{1}, k_{2},u}}^{\mathrm{H}}\mathbf{y}_u\right|^2\right]^2.
\end{aligned}
\end{equation}

To directly tackle \eqref{Neg18} with respect to $\mbX$, we write $\mby_{u}$ as $\mby_{u}=\mbX^{\top}\mbf_{u}$, where $\mbf_{u}=\left[\begin{array}{llll}
1 &  e^{-\textrm{j} 2 \pi \frac{u}{N}} & \cdots &  e^{-\textrm{j} 2 \pi \frac{(N-1)u}{N}}
\end{array}\right]^\top$ is the DFT vector. Then, \eqref{Neg18} becomes
\begin{equation}
\label{Neg20}
\begin{aligned}
&\underset{\mbx_m\in\Omega^{N}}{\textrm{minimize}} \sum_{k_{1}=1}^{K_{1}} \sum_{k_{2}=1}^{K_{2}} \sum_{u=0}^{N-1}\left[\widehat{P}{_{k_{1}, k_{2}, u}}-\left| \boldsymbol{\upalpha}_{_{k_{1}, k_{2},u}}^{\mathrm{H}} \mbX^{\top}\mbf_{u}\right|^2\right]^2,
\end{aligned}
\end{equation}
Expanding the objective $ \mathcal{P}=\left[\widehat{P}{_{k_{1}, k_{2}, u}}-\left| \boldsymbol{\upalpha}_{_{k_{1}, k_{2},u}}^{\mathrm{H}} \mbX^{\top}\mbf_{u}\right|^2\right]^2$, we obtain a quartic formulation is
\begin{equation}
\label{Neg21}
\begin{aligned}
\mathcal{P}= \mbf^{\mathrm{H}}_{u}\mbX^{\star}\boldsymbol{\upalpha}_{_{k_{1}, k_{2}, u}}\boldsymbol{\upalpha}^{\mathrm{H}}_{_{k_{1}, k_{2}, u}}\mbX^{\top}\mbf_{u}\mbf^{\mathrm{H}}_{u}\mbX^{\star}\boldsymbol{\upalpha}_{_{k_{1}, k_{2}, u}}\boldsymbol{\upalpha}^{\mathrm{H}}_{_{k_{1}, k_{2}, u}}\mbX^{\top}\mbf_{u}-2 \widehat{P}{_{k_{1}, k_{2},u}} \mbf^{\mathrm{H}}_{u}\mbX^{\star}\boldsymbol{\upalpha}_{_{k_{1}, k_{2}, u}}\boldsymbol{\upalpha}^{\mathrm{H}}_{_{k_{1}, k_{2}, u}}\mbX^{\top}\mbf_{u}+\widehat{P}^{2}{_{k_{1}, k_{2},u}}.
\end{aligned}
\end{equation}
Note that $\boldsymbol{\upalpha}^{\mathrm{H}}_{_{k_{1}, k_{2}, u}}\mbX^{\top}\mbf_{u}$ is scalar. Hence,
\begin{equation}
\label{Neg22}
\begin{aligned}
\mbf^{\mathrm{H}}_{u}\mbX^{\star}\boldsymbol{\upalpha}_{_{k_{1}, k_{2}, u}}\boldsymbol{\upalpha}^{\mathrm{H}}_{_{k_{1}, k_{2}, u}}\mbX^{\top}\mbf_{u}=\operatorname{vec}^{\top}\left(\mbf^{\mathrm{H}}_{u}\mbX^{\star}\boldsymbol{\upalpha}_{_{k_{1}, k_{2},u}}\right)\operatorname{vec}\left(\boldsymbol{\upalpha}^{\mathrm{H}}_{_{k_{1}, k_{2}, u}}\mbX^{\top}\mbf_{u}\right),
\end{aligned}  
\end{equation}\normalsize
where according to the identities of vectorization operator\cite{van1996matrix}, we have 
$\operatorname{vec}\left(\mbf^{\mathrm{H}}_{u}\mbX^{\star}\boldsymbol{\upalpha}_{_{k_{1}, k_{2},u}}\right)= \mbf^{\mathrm{H}}_{u}\operatorname{vec}\left(\mbX^{\star}\boldsymbol{\upalpha}_{_{k_{1}, k_{2},u}}\right)
=\mbf^{\mathrm{H}}_{u}\left(\boldsymbol{\upalpha}^{\top}_{_{k_{1},k_{2},u}}\otimes\mbI_{N}\right)\operatorname{vec}\left(\mbX^{\star}\right)$, and $\operatorname{vec}\left(\boldsymbol{\upalpha}^{\mathrm{H}}_{_{k_{1}, k_{2}, u}}\mbX^{\top}\mbf_{u}\right)= \boldsymbol{\upalpha}^{\mathrm{H}}_{_{k_{1}, k_{2}, u}}\operatorname{vec}\left(\mbX^{\top}\mbf_{u}\right)
= \boldsymbol{\upalpha}^{\mathrm{H}}_{_{k_{1}, k_{2}, u}}\left(\mbf^{\top}_{u}\otimes\mbI_{M}\right)\operatorname{vec}\left(\mbX^{\top}\right)$. Consequently,
\begin{equation}
\label{Neg25}
\begin{aligned}
\mbf^{\mathrm{H}}_{u}\mbX^{\star}\boldsymbol{\upalpha}_{_{k_{1}, k_{2}, u}}\boldsymbol{\upalpha}^{\mathrm{H}}_{_{k_{1}, k_{2}, u}}\mbX^{\top}\mbf_{u}=\operatorname{vec}^{\top}\left(\mbX^{\star}\right)\left(\boldsymbol{\upalpha}_{_{k_{1},k_{2},u}}\otimes\mbI_{N}\right)\mbf^{\star}_{u}\boldsymbol{\upalpha}^{\mathrm{H}}_{_{k_{1}, k_{2},u}}\left(\mbf^{\top}_{u}\otimes\mbI_{M}\right)\operatorname{vec}\left(\mbX^{\top}\right)
\end{aligned}
\end{equation}\normalsize
Using the commutation matrix $\mbP$, i.e., $\operatorname{vec}\left(\mbX^{\top}\right)=\mbP \operatorname{vec}\left(\mbX\right)$ and the fact that $\operatorname{vec}^{\top}\left(\mbX^{\star}\right)=\operatorname{vec}^{\mathrm{H}}\left(\mbX\right)$, \eqref{Neg25} becomes $\mbf^{\mathrm{H}}_{u}\mbX^{\star}\boldsymbol{\upalpha}_{_{k_{1}, k_{2}, u}}\boldsymbol{\upalpha}^{\mathrm{H}}_{_{k_{1}, k_{2}, u}}\mbX^{\top}\mbf_{u}=
\operatorname{vec}^{\mathrm{H}}\left(\mbX\right)\mbG\operatorname{vec}\left(\mbX\right)$,
where $\mbG=\left(\boldsymbol{\upalpha}_{_{k_{1}, k_{2},u}}\otimes\mbI_{N}\right)\mbf^{\star}_{u}\boldsymbol{\upalpha}^{\mathrm{H}}_{_{k_{1}, k_{2}, u}}\left(\mbf^{\top}_{u}\otimes\mbI_{M}\right)\mbP$.
Thus, the objective of \eqref{Neg20} is reformulated to
\begin{equation}
\label{Neg27}
\begin{aligned}
\mathcal{P}=\operatorname{vec}^{\mathrm{H}}\left(\mbX\right)\left(\mathcal{G}\left(\mbX\right)-2 \widehat{P}{_{k_{1}, k_{2},u}}\mbG\right)\operatorname{vec}\left(\mbX\right)+\widehat{P}^{2}{_{k_{1}, k_{2},u}}
\end{aligned}  
\end{equation}
where $\mathcal{G}\left(\mbX\right)=\mbG\operatorname{vec}\left(\mbX\right)\operatorname{vec}^{\mathrm{H}}\left(\mbX\right)\mbG$. The beampattern matching problem is now cast as the following quartic matrix program (QMP):
\begin{equation}
\label{Neg28}
\begin{aligned}
\underset{\mbx_m\in\Omega^{N}}{\textrm{minimize}} ~\operatorname{vec}^{\mathrm{H}}\left(\mbX\right)\widehat{\mbG}\left(\mbX\right)\operatorname{vec}\left(\mbX\right),
\end{aligned}
\end{equation}
with $\widehat{\mbG}\left(\mbX\right)=\sum_{k_{1}=1}^{K_{1}} \sum_{k_{2}=1}^{K_{2}} \sum_{u=0}^{N -1}\left[\mathcal{G}\left(\mbX\right)-2 \widehat{P}{_{k_{1}, k_{2},u}}\mbG\right]$.\normalsize
\subsection{WISL Criterion for Unimodular Waveform Design}
\label{WISL}
Consider a collection of $M$ unimodular waveforms, each of a code length of $N$. 
The cross-correlation between the $m$-th and $m^{\prime}$-th waveforms of sequences is 
$r_{m m^{\prime}}(k) = \sum_{l=0}^{N-k-1}x_{m}(l)x^{\star}_{m^{\prime}}(l+k) = r^{\star}_{mm^{\prime}}(-k)$ \cite{he2009designing}. Denote $\tau_{mmk}=\left|r_{m m}(k)\right|^2$ and $\eta_{mm^{\prime}k}=\left|r_{m m^{\prime}}(k)\right|^2$, the WISL criterion of waveform $\mbX$ is \cite{he2009designing}
\begin{align}
\label{eq:3}
\mathcal{W}=\sum_{m=1}^M \sum_{\substack{k=-N+1\\k\neq 0}}^{N-1} \omega_k^2\eta_{mmk}+\sum_{m=1}^M \sum_{\substack{m^{\prime}=1m^{\prime}\neq m}}^M \sum_{\substack{k=-N+1}}^{N-1}\omega_k^2\eta_{mm^{\prime}k},
\end{align}\normalsize
where $\left\{\omega_{k}\right\}^{N}_{k=1}$ are weights. 

The unimodular waveform with good correlation properties is obtained by solving the following optimization problem:
\begin{equation}
\label{Neg29}
\begin{aligned}
&\underset{\mbx_{m}\in\Omega^{N}}{\textrm{minimize}} ~\mathcal{W}.
\end{aligned}
\end{equation}
Following \cite{li2017fast}, 
this WISL minimization reduces to 
\begin{equation}
\label{Neg30}
\begin{aligned}
&\underset{\mbx_{m}\in\Omega^{N}}{\textrm{minimize}} ~\sum_{k=1}^{2 N}\left\|\mathbf{X}^{\mathrm{H}}\left(\left(\boldsymbol{\upbeta}_k\boldsymbol{\upbeta}^{\mathrm{H}}_k\right) \odot \boldsymbol{\Gamma}\right) \mathbf{X}\right\|^2_{\mathrm{F}},
\end{aligned}
\end{equation}
where $\bGamma\in\mathbb{R}^{N\times N}$
is a Toeplitz matrix whose upper and lower
triangular parts are constructed by the weight $\left\{\omega_{k}\right\}^{N-1}_{k=0}$ and $\left\{\omega_{-k}\right\}^{N-1}_{k=1}$, respectively, i.e., 
\begin{equation}
\boldsymbol{\Gamma} \triangleq\begin{bsmallmatrix}
\omega_0 & \omega_1 & \cdots & \omega_{N-1} \\
\omega_{-1} & \omega_0 & \ddots & \vdots \\
\vdots & \ddots & \ddots & \omega_1 \\
\omega_{-N+1} & \cdots & \omega_{-1} & \omega_0
\end{bsmallmatrix},
\end{equation}
and $\boldsymbol{\upbeta}_{k}=\left[\begin{array}{llll}
1 &  e^{\textrm{j} 2 \pi \frac{(k-1)}{2N}} & \cdots &  e^{\textrm{j} 2 \pi \frac{(N-1)(k-1)}{2N}}
\end{array}\right]^\top$.

\subsection{Low-WISL Waveform Design as UQMP}
To tackle the WISL minimization problem with our proposed algorithm, which is a variant of the power iteration method, we change the objective to bring it in the form $\mbs^{\mathrm{H}} \mbR \mbs,~\mbs\in\mathbb{C}^{N},~\mbR\in\mathbb{R}^{N\times N}$ that is suitable for our algorithm steps. Substitute $\mbJ_{k}=\left(\boldsymbol{\upbeta}_k\boldsymbol{\upbeta}^{\mathrm{H}}_k\right) \odot \boldsymbol{\Gamma}$ in the objective as 
\begin{equation}
\label{Neg31}
\begin{aligned}
\left\|\mathbf{X}^{\mathrm{H}}\mbJ_{k} \mathbf{X}\right\|^2_{\mathrm{F}}&= \operatorname{Tr}\left(\mbX^{\mathrm{H}}\mbJ^{\mathrm{H}}_{k}\mbX\mbX^{\mathrm{H}}\mbJ_{k}\mbX\right),\\
&=\operatorname{vec}^{\top}\left(\mbX^{\top}\mbJ^{\star}_{k}\mbX^{\star}\right)\operatorname{vec}\left(\mbX^{\mathrm{H}}\mbJ_{k}\mbX\right),\\
&=\operatorname{vec}^{\mathrm{H}}\left(\mbX\right)\left(\mbI_{M}\otimes \mbX^{\top}\mbJ^{\star}_{k}\right)^{\top}\left(\mbI_{M}\otimes \mbX^{\mathrm{H}}\mbJ_{k}\right)\operatorname{vec}\left(\mbX\right),\\
&=\operatorname{vec}^{\mathrm{H}}\left(\mbX\right)\left(\mbI_{M}\otimes \mbJ^{\mathrm{H}}_{k}\mbX\mbX^{\mathrm{H}}\mbJ_{k}\right)\operatorname{vec}\left(\mbX\right).
\end{aligned}
\end{equation}\normalsize
Define $\mathcal{J}\left(\mbX\right)=\sum^{2N}_{k=1}\left(\mbI_{M}\otimes \mbJ^{\mathrm{H}}_{k}\mbX\mbX^{\mathrm{H}}\mbJ_{k}\right)=\mbI_{M}\otimes\left(\sum^{2N}_{k=1}\mbJ^{\mathrm{H}}_{k}\mbX\mbX^{\mathrm{H}}\mbJ_{k}\right)$. The WISL minimization problem is now recast as a UQMP as follows:
\begin{equation}
\label{Neg32}
\begin{aligned}
&\underset{\mbx_{m}\in\Omega^{N}}{\textrm{minimize}} ~\operatorname{vec}^{\mathrm{H}}\left(\mbX\right)\mathcal{J}\left(\mbX\right)\operatorname{vec}\left(\mbX\right).
\end{aligned}
\end{equation}

Now, both \eqref{Neg28} and \eqref{Neg32} share the same form and can be optimized together in a single optimization problem. Hence, we consider the following optimization problem that designs a unimodular waveform with a low-WISL that simultaneously satisfies beampattern matching requirements:
\begin{equation}
\label{Neg33}
\begin{aligned}
\underset{\mbx_{m}\in\Omega^{N}}{\textrm{minimize}}~ \gamma\mathcal{P}+(1-\gamma)\mathcal{W}.
\end{aligned}
\end{equation}
where $0\leq\gamma\leq 1$ is the Lagrangian multiplier. The resulting UQMP is
\begin{equation}
\label{Neg34}
\begin{aligned}
&\underset{\mbx_{m}\in\Omega^{N}}{\textrm{minimize}} ~\operatorname{vec}^{\mathrm{H}}\left(\mbX\right)\left(\gamma \widehat{\mbG}\left(\mbX\right)+(1-\gamma)\mathcal{J}\left(\mbX\right)\right)\operatorname{vec}\left(\mbX\right).
\end{aligned}
\end{equation}
\section{Proposed Algorithm}
\label{sec:CyPMLI}
Our approach to solve the low-WISL waveform design problem \eqref{Neg34} is to cast it as a UBQMP and then tackle it using the CyPMLI algorithm. Define $\mbR\left(\mbX\right)=\left(\gamma \widehat{\mbG}\left(\mbX\right)+(1-\gamma)\mathcal{J}\left(\mbX\right)\right)$. To transform \eqref{Neg34} into two quadratic optimization subproblems, we define two variables $\operatorname{vec}\left(\mbX_1\right)$ and $\operatorname{vec}\left(\mbX_2\right)$. It is also interesting to observe that if either $\mbX_1$ or $\mbX_2$   are fixed, solving \eqref{Neg34}
 with respect to the other variable can be done
via a unimodular quadratic programming (UQP) formulation:
\begin{equation}
\label{Neg35}
\begin{aligned}
\underset{\operatorname{vec}\left(\mbX_j\right)\in\Omega^{N M}}{\textrm{minimize}}\quad \operatorname{vec}^{\mathrm{H}}\left(\mbX_{j}\right) \mbR\left(\mbX_{i}\right) \operatorname{vec}\left(\mbX_{j}\right), \quad i\neq j \in \left\{1,2\right\}.
\end{aligned}
\end{equation}
Note that if either $\mbX_1$ or $\mbX_2$ are fixed, minimizing the objective with respect to the other variable is achieved via UQP \cite{eamaz2023cypmli,esmaeilbeig2022joint}.
To ensure the convergence of $\mbX_1$ and $\mbX_2$ to the same waveform, a connection needs to be established between them in the objective. 
Adding the Frobenius norm error between $\mbX_{1}$ and $\mbX_{2}$ as a \emph{penalty} with the Lagrangian multiplier to (\ref{Neg35}), we have the following \emph{regularized Lagrangian} problem:
\begin{equation}
\label{Neg37}
\begin{aligned}
\underset{\operatorname{vec}\left(\mbX_j\right)\in\Omega^{N M}}{\textrm{minimize}}~ \operatorname{vec}^{\mathrm{H}}\left(\mbX_{j}\right) \mbR\left(\mbX_{i}\right) \operatorname{vec}\left(\mbX_{j}\right)+\rho \left\|\mbX_{i}-\mbX_{j}\right\|^{2}_{\mathrm{F}},
\end{aligned}
\end{equation}
where $\rho$ is the Lagrangian multiplier. 
The penalty $\left\|\mbX_{i} -\mbX_{j}\right\|_{\mathrm{F}}^{2}$ is also a quadratic function with respect to $\mbX_{j}$. Consequently, the UBQMP formulation for \eqref{Neg34} is given by below
\begin{equation}
\label{Neg38}
\begin{aligned}
\underset{\operatorname{vec}\left(\mbX_j\right)\in\Omega^{N M}}{\textrm{minimize}}
\begin{psmallmatrix}
\operatorname{vec}\left(\mbX_{j}\right)\\1\end{psmallmatrix}^{\mathrm{H}}
\underbrace{\begin{psmallmatrix}
\mbR(\mbX_{i})&
-\rho\operatorname{vec}\left(\mbX_{i}\right) \\ 
-\rho \operatorname{vec}^{\mathrm{H}}\left(\mbX_{i}\right)&
2\rho N M\end{psmallmatrix}}_{\breve{\mbR}(\mbX_{i})}
\begin{psmallmatrix}
\operatorname{vec}\left(\mbX_{j}\right)\\1\end{psmallmatrix},
\end{aligned}
\end{equation}
To employ CyPMLI, we need to change the problem to a maximization problem using the \emph{diagonal loading process}. Denote the maximum eigenvalue of $\breve{\mbR}\left(\mbX_{i}\right)$ by $\lambda_{m}$, where $\lambda_{m}\mbI\succeq \breve{\mbR}\left(\mbX_{i}\right)$.
Thus, $\widehat{\mbR}\left(\mbX_{i}\right) = \lambda_{m} \mbI - \mbR\left(\mbX_{i}\right)$ is positive definite\cite{eamaz2023cypmli}. Note that a diagonal loading with $\lambda_{m}\mbI$ has no effect on the solution of (\ref{Neg38}) due to the fact that $\left\|\mbX\right\|^{2}_{\mathrm{F}}=N M$ and $\operatorname{vec}^{\mathrm{H}}\left(\mbX_{j}\right) \widehat{\mbR}\left(\mbX_{i}\right) \operatorname{vec}\left(\mbX_{j}\right)=\lambda_{m}N M-\operatorname{vec}^{\mathrm{H}}\left(\mbX_{j}\right) \mbR\left(\mbX_{i}\right) \operatorname{vec}\left(\mbX_{j}\right)$. Therefore, we have
\begin{equation}
\label{Neg39}
\begin{aligned}
\underset{\operatorname{vec}\left(\mbX_j\right)\in\Omega^{N M}}{\textrm{maximize}}
\begin{psmallmatrix} 
\operatorname{vec}\left(\mbX_{j}\right)\\1\end{psmallmatrix}^{\mathrm{H}}
\underbrace{\begin{psmallmatrix}
\widehat{\mbR}(\mbX_{i}) &
\rho\operatorname{vec}\left(\mbX_{i}\right)\\
\rho\operatorname{vec}^{\mathrm{H}}\left(\mbX_{i}\right)&
\widehat{\rho}\end{psmallmatrix}}_{\mathcal{R}(\mbX_{i})}
\begin{psmallmatrix} 
\operatorname{vec}\left(\mbX_{j}\right)\\1\end{psmallmatrix},
\end{aligned}
\end{equation}\normalsize
where $\widehat{\rho}=\lambda_{m}-2\rho N M$. The desired matrix $\mbX_{j}$ of \eqref{Neg39} is readily evaluated by PMLI in each iteration as $\boldsymbol{\upnu}^{(t+1)}=e^{\textrm{j} \operatorname{arg}\left(\mathcal{R}(\mbX_{i})
\boldsymbol{\upnu}^{(t)}\right)}$ \cite{eamaz2023cypmli}, where $\boldsymbol{\upnu}=\left(\operatorname{vec}^{\top}\left(\mbX_{j}\right)~1\right)^{\top}$. This update process can be simplified as
\begin{equation}
\label{NegAr}
\operatorname{vec}\left(\mbX^{(t+1)}_{j}\right)=e^{\textrm{j} \operatorname{arg}\left(\widehat{\mbR}\left(\mbX^{(t)}_{i}\right)\operatorname{vec}\left(\mbX^{(t)}_{j}\right)+\rho\operatorname{vec}\left(\mbX^{(t)}_{i}\right)\right)}.
\end{equation}
The update process (\ref{NegAr}) requires only a simple matrix vector multiplication while leveraging information about previous update $\mbs^{(t)}_{i}$ through the momentum term. This update process resembles that of the gradient descent projection with a \emph{heavy ball momentum}, where in each iteration, the solution information from the previous step is incorporated using a momentum term \cite{loizou2020momentum}.
Such power method-like iterations are already shown to be convergent in terms of the signals\cite{soltanalian2014designing,eamaz2023marli}, implying that $\mbX_{1}$ and $\mbX_{2}$ will be converging to each other as well.
\begin{figure}[t]
	\centering
	\subfloat[]
		{\includegraphics[width=0.5\columnwidth]{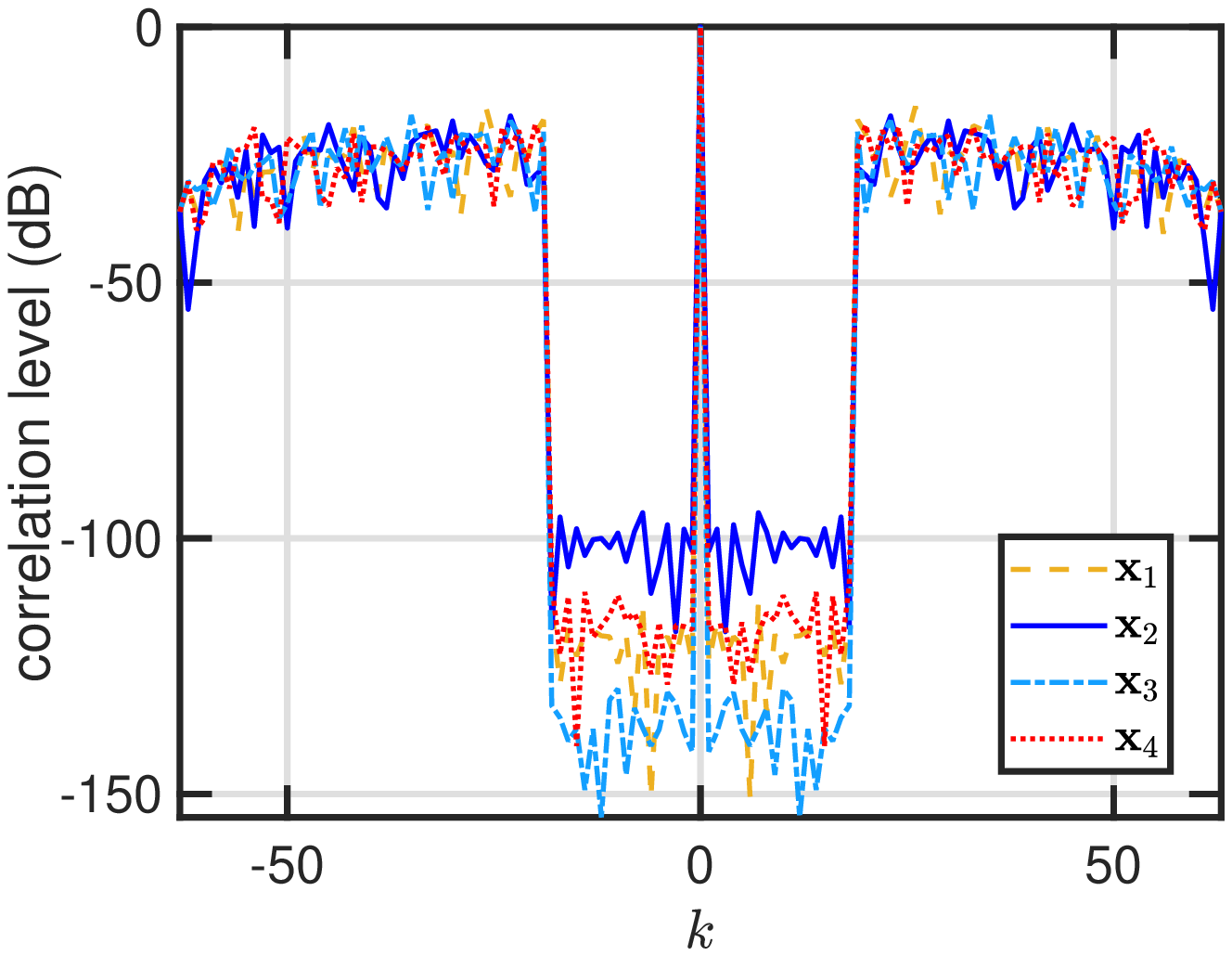}}
	\subfloat[]
		{\includegraphics[width=0.5\columnwidth]{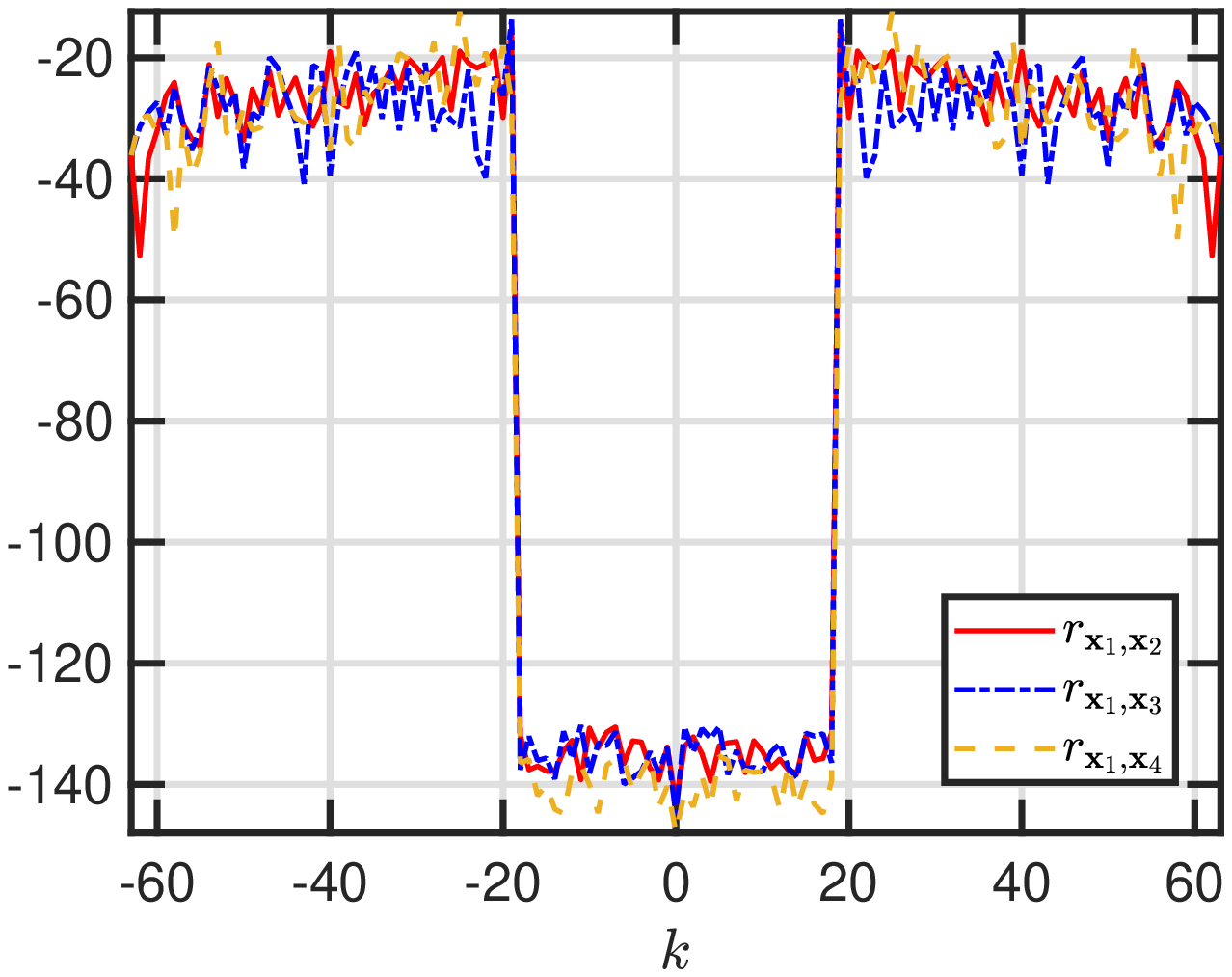}}
    \caption{(a) The correlation level of the designed waveform with sequence length $N=64$ for antenna array with $M=4$ elements. (b) The cross-correlation between $\mbx_{1}$ and other sequences in the designed waveform; i.e. $\mbx_{2}$, $\mbx_{3}$ and $\mbx_{4}$ with the same values of $N$ and $M$ as in (a).
 }
	\label{figure_1}
\end{figure}

\begin{figure}[t]
	\centering
	\subfloat[]
		{\includegraphics[width=0.5\columnwidth]{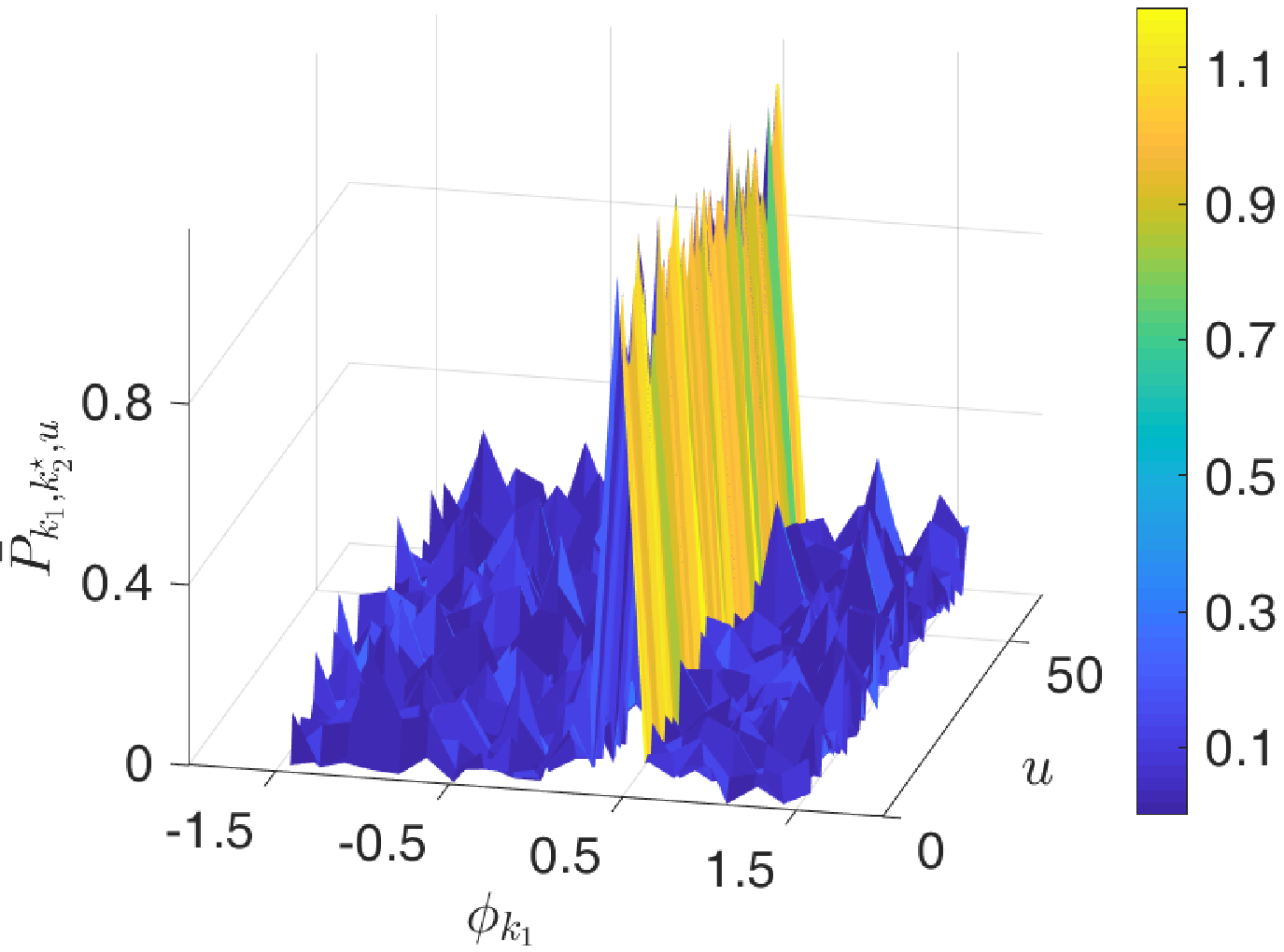}}
	\subfloat[]
		{\includegraphics[width=0.5\columnwidth]{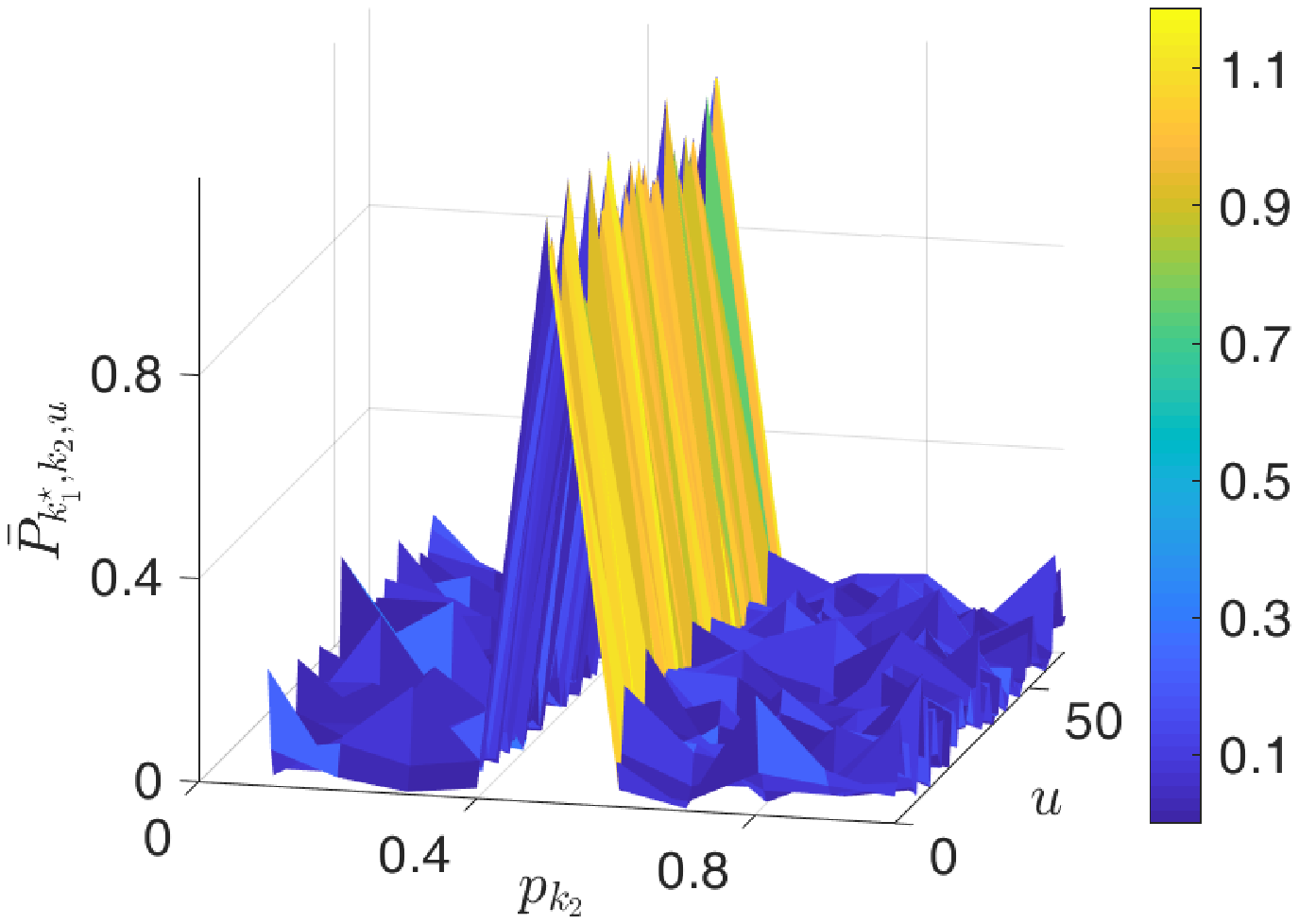}}
    \caption{The obtained near-field beampattern with respect to (a) $-\frac{\pi}{2}\leq\phi_{k_{1}}\leq\frac{\pi}{2}$ and $u$ for fixed $k_{1}=k_{1}^{\star}$, and (b) $0<p_{k_{2}}\leq 1$ and $u$ for fixed $k_{2}=k_{2}^{\star}$. In all cases, $N=64$ and $M=4$.
    }
	\label{figure_2}
\end{figure}

\section{Numerical Experiments}
\label{sec:num_exp}
We numerically evaluated the efficacy of our approach. We used the following settings for our experiments: the number of array elements $M=4$, the carrier frequency of the transmitted signal $f_{c}=1~\mathrm{GHz}$, the bandwidth $B=200~\mathrm{MHz}$, and the number of symbols $N=64$. The inter-element spacing is $d=c_{0}/(2(f_{c}+B/2))$ (half wavelength of the highest in-band frequency) to avoid grating lobes. The DoA (normalized range) domain set to $-\frac{\pi}{2}\leq\phi\leq\frac{\pi}{2}$ ($0<p\leq 1$) was discretized with $K_{1}=20$ ($K_{2}=10$) grid points.

The CyPMLI parameters were $\rho=2$ and $\gamma=0.5$. We updated the value of $\lambda_{m}$ according to \cite[Theorem 1]{eamaz2023cypmli}. Fig.~\ref{figure_1}a shows that the resulting waveform achieves a satisfactory correlation level. Further, the designed sequences exhibit a good cross-correlation property with each other (Fig.~\ref{figure_1}b). 
Assume that the desired beampattern is $1$ at the indices $k_{1}^{\star}$ and $k_{2}^{\star}$ and $0$ elsewhere for all $u$. Fig.~\ref{figure_2}a displays the (near-field) beampattern obtained for the angular span of $-\frac{\pi}{2}\leq\phi_{k_{1}}\leq\frac{\pi}{2}$ and discrete frequency $u$ with fixed $k_{1}=k_{1}^{\star}$. On the other hand, Fig.~\ref{figure_2}b shows the beampattern as a function of range $0<p_{k_{2}}\leq 1$ and $u$ with fixed $k_{2}=k_{2}^{\star}$.  In both cases, CyPMLI maintains good input correlation properties as shown in Fig.~\ref{figure_1} while 
obtaining the desired beampattern with a small negligible error.
\section{Summary}
\label{sec:summ}
THz automotive radars are expected to provide near-optical resolution very close to that of lidars. For the ultrashort range operation, near-field propagation needs to be considered in the waveform design for these systems. Our CyPMLI approach to obtain low-WISL unimodular waveforms suggests a way forward to realize the range-dependent beampattern in near-field. Future investigations include a comprehensive evaluation of this method.

\section*{Acknowledgement}
A.E., F.Y., and M.S. acknowledge partial support via National
Science Foundation Grant ECCS-1809225.

\bibliographystyle{IEEEtran}
\bibliography{references}

\end{document}